\documentclass[aps,twocolumn,floats,prx,nofootinbib,superscriptaddress]{revtex4-1}
\usepackage{amsmath, amsfonts}
\usepackage{graphicx}
\usepackage{tikz}
\usepackage{url}
\usepackage{bbm}
\usepackage{textgreek,bm}

\newcommand{\clim}{c_{\rm{lim}}}
\newcommand{\STA}{\text{STA}}
\newcommand{\Wmsr}{Wm$^{-2}$sr$^{-1}$ }
\newcommand{\E}{\mathbb{E}}

\begin{document}

\title{Context dependent adaptation in a neural computation}

\author{Charles J. Edelson\footnote}
\email{Contact Author: ce3721@princeton.edu \\ Current Address: Princeton University, Princeton, NJ 08544}
\affiliation{Department of Physics, Indiana University, Bloomington, IN 47405}
\author{Sima Setayeshgar}
\affiliation{Department of Physics, Indiana University, Bloomington, IN 47405}
\author{William Bialek}
\affiliation{Joseph Henry Laboratories of Physics, Princeton University, Princeton, NJ 08544}
\author{Rob R. de Ruyter van Steveninck}
\affiliation{Department of Physics, Indiana University, Bloomington, IN 47405}

\date{\today}

\begin{abstract}
Brains adapt to the statistical structure of their input. In the visual system, local light intensities change rapidly, the variance of the intensity changes more slowly, and the dynamic range of contrast itself changes more slowly still. We use a motion--sensitive neuron in the fly visual system to probe this hierarchy of adaptation phenomena, delivering naturalistic stimuli that have been simplified to have a clear separation of time scales. We show that the neural response to visual motion depends on contrast, and this dependence itself varies with context. Using the spike--triggered average velocity trajectory as a response measure, we find that context dependence is confined to a low--dimensional space, with a single dominant dimension.  Across a wide range of conditions this adaptation serves to match the integration time to the mean interval between spikes, reducing redundancy.
\end{abstract}

\maketitle

\section{Introduction}

Adaptation is a central feature of living systems, but ``adaptation'' means many things.   The dynamics of evolution results (in part) in adaptation by natural selection \cite{lenski2017}, and we describe features of the organism that enhance fitness as being adaptive.  This form of adaptation has been observed over tens of thousands of generations in {\em E.~coli}, but the same bacteria adapt their chemotactic behavior to constant background concentrations over the course of seconds \cite{block+al1983}.  We humans also adapt to constant sensory stimuli, becoming largely insensitive to absolute signals while retaining sensitivity to small variations around the constant background.  Correlates of such sensory adaptation were first characterized in single neurons one hundred years ago \cite{adrian1928}.  We now understand that sensory adaptation is more than simply subtracting a constant background \cite{wark2007sensory,weber+al_2019}.

The natural world is complex, with its statistical characteristics shifting across both space and time. Signals can change gradually, as with the subtle shift of light intensity when summer turns to fall, or abruptly, as with the dramatic shift experienced when moving from a dense forest to an open field. Despite this complexity and variability, animals go about their daily lives, reliably responding to the signals that matter most for their survival. This simple observation helped inspire Barlow's efficient coding hypothesis, which posits that sensory systems efficiently encode their natural stimuli  \cite{barlow_59,barlow1961possible}. The key insight  is that efficient coding requires sensory systems to be contextually aware and responsive to the \textit{distribution} of natural stimuli. This shift toward a distributional and information-theoretic perspective provided a powerful framework for studying sensory information encoding, with visual systems serving as an effective testing ground for these ideas \cite{laughlin_1981, smirnakis_adaptation_1997,egelhaaf2002vision, bialek2006efficient, wark2007sensory, kohn2007visual, bialek2012,webster2015visual, palmer2015predictive, tkavcik2016information,weber+al_2019}.\footnote{The idea of matching to the distribution of inputs is applied most often to strategies for coding.  But when signals are noisy or ambiguous, strategies for inferring things of interest to the organism also should adapt---not just neural coding but also neural computation should match the input distribution.  The computation of motion in the visual system provided an early modern example of this idea \cite{potters1994statistical,weiss+al2002}, which has precursors reaching back to Helmholtz \cite{barlow_1990}. These ideas expanded to a broader view of perception as Bayesian inference \cite{knill+richards_1996}.}

Fly visual systems, in particular, provide a valuable model for studying neural information encoding. The fly's optical lobes are organized into distinct layers, each of which has been thoroughly researched, with recent studies shedding light on the complete \textit{Drosophila} connectome \cite{flywire1}. The discovery of these well-organized structures has prompted questions regarding their functions, resulting in numerous efforts to understand how each component contributes to the fly's ability to see and interact with the world \cite{borst2023flies}. While substantial progress has been made overall, studies focusing on fly visual motion perception have been particularly successful. This can be at least partially attributed to the early discovery of the lobula plate tangential cells (LPTC), a collection of wide-field motion sensitive neurons located on their eponymous neural structure, the lobula plate \cite{bishop1966two}. These cells are directionally selective, and appear to be an important part of the fly's flight control system, with experimental evidence for LPTC involvement in several key flight behaviors \cite{pierantoni1976look, hausen1984lobula, krapp2008estimation, huston2008visuomotor, ullrich2014influence, wei2020diversity}. 

Given the importance of flight to flies, it is perhaps unsurprising that LPTCs maintain robust velocity responses even in the presence of changing scene statistics. For example, H1, a spiking horizontal wide-field motion sensitive LPTC, adapts to changes in light intensity, contrast, and velocity, and in some cases to the statistics of these variations \cite{maddess1985adaptation, de1986adaptation, brenner2000adaptive, fairhall2001efficiency, reisenman2003adaptation, flanagin2006dynamic}. This adaptive behavior, coupled with its experimental accessibility and role in flight control, has made H1 an attractive system for studying how neurons encode information in dynamic natural environments \cite{spikesbook,brenner_synergy_2000, brenner2000adaptive, fairhall2001efficiency, van2002timing, nemenman2008neural}. Adding to this experimental appeal is that visual stimuli are easy to control in a lab environment, allowing for precise manipulation of stimuli and the distributions out of which these stimuli are drawn. This comes with two major advantages. The first is the ability to ask very precisely \textit{what} causes a spike \cite{spikesbook, chichilnisky2001simple}. The second is the ability to repeat stimuli and measure the reliability of H1's response \cite{de1997reproducibility, van2002timing, nemenman2008neural}. When combined, these methods can provide powerful results about the optimality and efficiency of a neural code. As an example,  the dynamic range of H1's input/output  relation is scaled to the width of the input velocity distribution, maximizing information transmission \cite{brenner2000adaptive}.

In this paper, we expand upon these results and attempt to characterize how H1 adapts to both dynamic and distributional changes in the contrast of visual stimuli.  Quantitatively, we characterize  H1's response to motion by the average trajectory of velocity vs~time in the neighborhood of a spike---the spike triggered average \cite{deboer+kuyper_68} or STA.  We confirm that this response to motion depends strongly on the variance of light intensity in a scene,  or contrast, and demonstrate that this dependence itself is different when contrast variations are drawn from different distributions.  That is, the contrast variation of H1's motion response  depends on the range of contrast values that the fly has seen recently. These adaptations span a low rank subspace of possible STAs, with the majority of the variance occurring along a single stimulus dimension. This context dependence persists even when the contrast samples its distribution with correlation times across a broad range of scale,  $\tau_c = 0.25 -15\,{\rm s}$.  A surprising consequence of these adaptive dynamics is that the time scale for integration of velocity information tracks the mean interval between action potentials.  This reduces redundancy and shows that the system actively avoids a ``rate coding'' regime in which local velocities would be represented by multiple spikes \cite{spikesbook}.

\section{Stimuli and responses}

Visual signals in the real world have many layers of variation.  To search for different forms of adaptation it is convenient to create signals that capture some of these natural statistics but allow a clean separation between successive layers.  Here we describe our approach to this problem, admitting at the outset that many other approaches are possible.  We also face the issue of characterizing the neural responses to these complex signals, and defining the signatures of adaptation and context dependence.

\subsection{Intensity, contrast, and context}
\label{sec:control_contrast}

A visual stimulus is defined as a time-dependent two-dimensional intensity field, $I(x, y, t)$, where $x$ and $y$ are angular coordinates on the retina.   We define the instantaneous contrast as the fractional root-mean-square variations in intensity across space,
\begin{equation}
    \label{eq:rms_contrast}
    c(t) = \text{C}[I(x, y, t)] \equiv \frac{\sqrt{\text{Var}_{x, y} [ I(x, y, t)]}}{\E_{x, y}[I(x, y, t)]},
\end{equation}
where $\E_{x,y}[\cdot]$ and $\rm{Var}_{x,y}[\cdot]$ are the expectation and variance over space.  Because motion is a dominant source of intensity variation, fractional variance over time at a single point is very similar to the spatial contrast that we define here.

With these considerations in mind, to design a stimulus with a specific contrast distribution and mean light intensity, we start with a static pattern, $F(x, y)$, with zero mean and unit variance.    A horizontal velocity trajectory, $v(t)$, is generated and integrated to produce a motion trajectory, 
\begin{equation}
x(t) = x_0 + \int_0^t d\tau \, v(\tau )
\end{equation}
$x(t)$. The dynamic contrast stimulus is then constructed by translating the pattern along this trajectory,
\begin{equation}
    \label{eq:contrast_stimulus}
    I(x, y, t) = \bar{I} \left[ 1 + c(t) \cdot F(x(t),y) \right],
\end{equation}
where $\bar{I} > 0$ and $c(t) \in [-1, 1]$. The first parameter $\bar{I}$ sets the mean light intensity of the scene, and $c(t)$ is contrast as defined in Eq~(\ref{eq:rms_contrast}). 

Zero contrast implies that $I(x, y, t)$ is almost everywhere constant and hence visual inference of the underlying velocity $v(t)$ is impossible.  Controlling  contrast gives one method of adjusting the signal-to-noise ratio (SNR) for any visual inference. Similarly, the mean light intensity controls the SNR  setting the  noise level due to random arrival of photons, as well as other sources of receptor noise.

We could choose $F(x,y)$ directly from the natural environment, but we would like a bit more control. 
Motivated by the observation of scale invariance in natural scenes \cite{burton1987color, field1987relations, ruderman_statistics_1994}, we start in Fourier space with
\begin{equation}
    \tilde{F}_{\rm sns}(k_x, k_y) = \frac{A_0e^{i \phi (k_x, k_y)}}{\sqrt{k_x^2 + k_y^2}},
\end{equation}
where $(k_x,k_y)$ are the spatial frequencies  conjugate to $(x,y)$, ?and ``sns" refers to our construction of synthetic natural scenes.  The overall amplitude $A_0$ is arbitrary and will drop out; we choose the phases $\phi (k_x, k_y)$ independently at each spatial frequency from the uniform distribution on $0 \leq \phi < 2\pi$.  We then transform back into real space,\footnote{As with $A_0$, because we ultimately binarize the images we don't need to keep track of all the proportionality constants.  Note also that since $F$ is real we can set only half of the phases; the other half are determined by $\phi (-k_x,- k_y) = -\phi (k_x, k_y)$.}
\begin{equation}
F_{\rm sns} (x,y) \propto \sum_{k_x , k_y}  \tilde{F}_{\rm sns}(k_x, k_y) \exp\left[ +i(k_x x + k_y y)\right],
\end{equation}
binarize depending on whether the local value is above or below the spatial mean, and finally rescale to be sure that the resulting $F(x,y)$ has zero mean and unit variance.  Notice that because we have only two possible values for $F$ the intensity in Eq~(\ref{eq:contrast_stimulus}) is guaranteed to be positive.

This slightly circuitous method of generating scenes allows us to vary the contrast while holding the spatial correlations fixed.  The binary structure means that we do not have to worry about tails of the distribution exceeding the dynamic range of the display, and the whole construction can be done at arbitrary spatial resolution.  The pattern $F(x,y)$ naturally obeys periodic boundary conditions, allowing for smooth wrapping along display edges.

We emphasize that the stimulus $I(x,y,t)$ varies for three reasons.  First, the movements $v(t)$ can produce rapid variations in light intensity. In practice will take $v(t)$ to be Gaussian white noise, so that the displacements $x(t)$ are diffusive.  In each $\Delta \tau = 2\,{\rm ms}$ time step of our digital display (see below), the velocity variance is $\langle v^2\rangle = \left(37\, {\rm deg/s}\right)^2$, so the pattern diffuses
\begin{eqnarray}
\langle | x(t) - x(t')|^2\rangle &=& 2 D| t-t'|\\
D &=& \langle v^2\rangle\Delta\tau/2 = 1.37\, {\rm deg}^2/{\rm s} .
\end{eqnarray}

Second, the stimulus varies because we allow the contrast $c(t)$ to vary.   Specifically, we choose $c(t)$ uniformly $-c_{\rm lim} \leq c < c_{\rm lim}$, where from Eq~(\ref{eq:contrast_stimulus}) we see that negative contrast corresponds to a black/white reversal of image intensities.  We introduce temporal correlations by introducing a Gaussian random function $u(t)$ with the correlation function
\begin{equation}
\langle u(t) u(t') \rangle = {1\over 3} c_{\rm lim}^2\exp\left( - |t - t'| /\tau_c \right) ,
\end{equation}
and then make a moment-by-moment nonlinear transformation $u(t) \rightarrow c(t)$ to insure that the distribution of $c$ is uniform.  In all the experiments discussed here we have $\tau_c = 500\,{\rm ms}$.  Importantly we will see that this time scale is longer than the integration time of the neural response to motion.

Finally, the stimuli vary because the dynamic range of contrast variations $c_{\rm lim}$ can be changed.  We do this on the longest time scale $\sim 30\,{\rm min}$, essentially doing experiments in successive blocks with different value of $c_{\rm lim}$.  In this sense it is natural to describe the setting of $c_{\rm lim}$ as the context for contrast changes.

\subsection{Inputs, outputs, and correlations}
\label{sec:in_out_and_corr}

We will analyze the responses of the motion sensitive neuron H1 in the fly visual system.  This is a spiking neuron, as with most cells in the brain, which means that its electrical activity consists of a sequence of discrete, identical voltage pulses, each  with roughly millisecond duration.\footnote{See, for example, Figure 1.2 in Ref \cite{spikesbook}.}  These action potentials or spikes thus can be defined  by their arrival times $\{t_i \}$, and the overall output or response of the neuron is
\begin{equation}
z(t) = \sum_{i=1}^N \delta (t - t_i) ,
\end{equation}
where the experiment is done in some large interval of time $T$ where we observe $N$ spikes.  Given the inputs are drawn from some statistical distribution or ensemble, we can define an average rate of spikes
\begin{equation}
\bar r =  \lim_{T\rightarrow\infty} {1\over T}\int_0^T dt\, z(t) .
\end{equation}

It is an old idea that we can characterize the dynamics of complex, nonlinear systems by estimating correlation functions between inputs and outputs \cite{wiener_58}.  Since H1 is sensitive to visual motion, we can think of the input as the (zero mean) velocity $v(t)$, and so a natural correlation function is
\begin{eqnarray}
C_{zv}(\tau ) &=& \langle z(t) v(t+\tau )\rangle \\
&=& \lim_{T\rightarrow\infty} {1\over T}\int_0^T dt\,\sum_{i=1}^N \delta (t - t_i) v(t+\tau )\\
&=& \bar r \langle v(t_i + \tau )\rangle ,
\end{eqnarray}
which we recognize as being proportional to the mean input in the neighborhood of a spike.  It is useful to isolate this ``spike--triggered average''
\begin{equation}
{\rm STA}(\tau ) = \langle v(t_i + \tau )\rangle.
\label{STA1}
\end{equation}

\begin{figure}
    \centering
    \includegraphics[width=\linewidth]{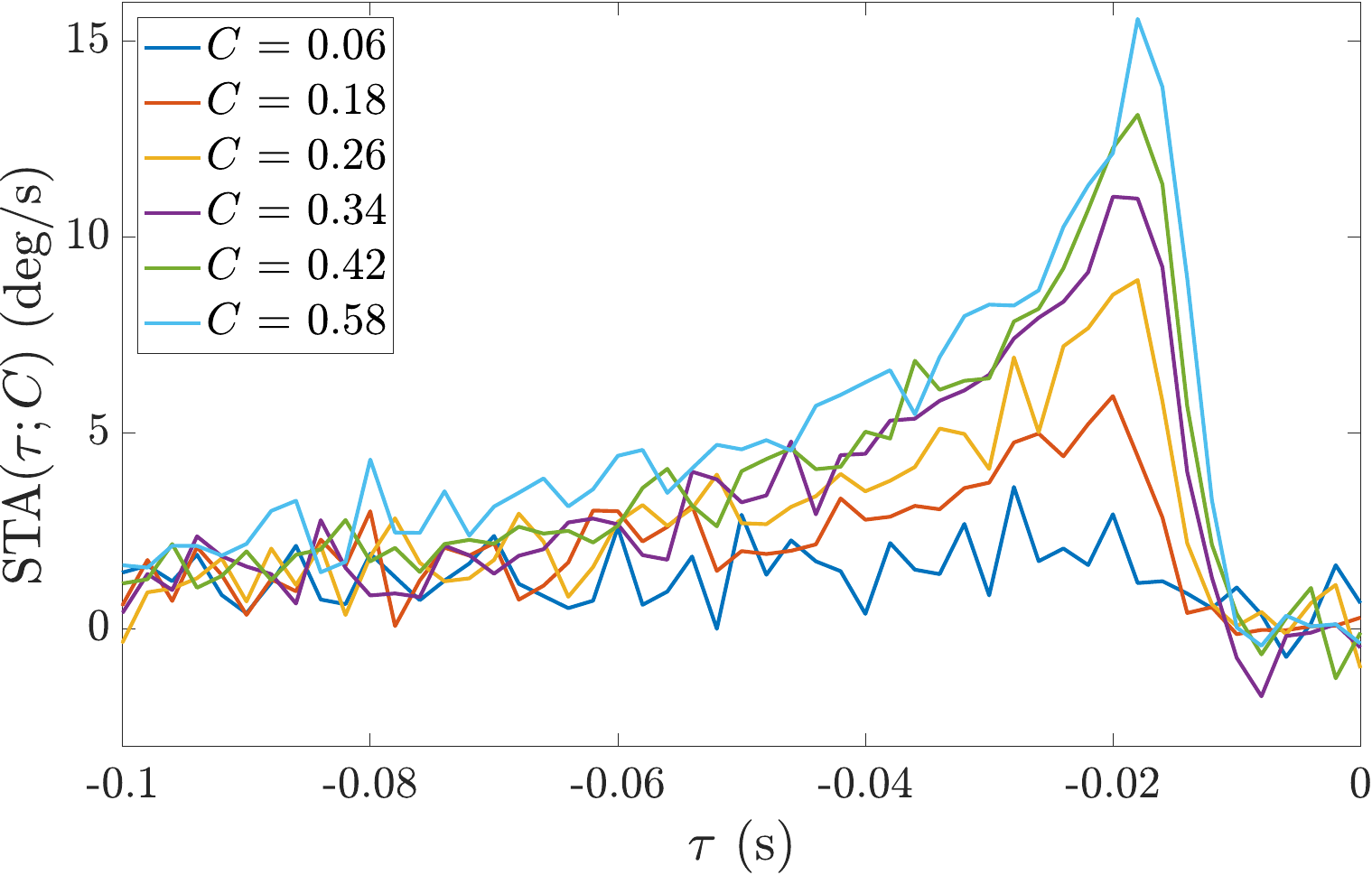}
    \caption{Contrast dependent spike triggered averages, ${\rm STA}(\tau; C)$ from Eq~(\ref{eq:STAC}). From an experiment with the dynamic range contrasts $c_{\rm lim} = 0.6$.   Curves are the mean across a $30\,{\rm min}$ experiment, and high frequency fluctuations give the scale of the measurement errors. 
    \label{fig:stas_clim_60}}
\end{figure}

Notice that if there are no correlations in $v(t)$ itself, as with the white noise signals used here, responses can be correlated only with inputs in the past, so that $C_{zv}(\tau >0) = 0$.  This suggests that the correlation function is effectively a response function, and this can be made precise.  Concretely, if we imagine the neuron being driven by the input $v(t)$ many times, we can average over these multiple ``trials'' to compute the time--dependent rate or probability per unit time of observing a spike
\begin{equation}
r(t) = {\bigg \langle} \sum_{i=1}^N \delta (t - t_i) {\bigg\rangle }_v ,
\end{equation}
where the subscript reminds us that the input trajectory is given.  If the neuron is responding not to arbitrary variation in $v(t)$ but only to some filtered version of this input then we should have
\begin{equation}
r(t) = \bar r G\left[ \int d\tau f(\tau ) v(t-\tau )\right ],
\end{equation}
where $G[\cdot]$ is some nonlinear function.  When $v(t)$ is Gaussian white noise it can be shown that the filter that characterizes the neural response is proportional to the spike--triggered average  \cite{deboer+kuyper_68},
\begin{equation}
 {\rm STA}(\tau ) = 2 D \langle G'\rangle f(-\tau ) .
\end{equation}
Thus we can think of the STA as a short snippet of velocity vs time that is associated with each spike, or as the direction in stimulus space to which the cell is most responsive.  Consistent with previous work \cite{maddess1985adaptation, de1986adaptation, de1995reliability} we will see that the time scales of the STA are tens of milliseconds (see, for example, Fig \ref{fig:stas_clim_60}).

Since we characterize neural response to motion by computing a correlation function, we can ask about the role of other variables by conditioning the average.  Specifically we can define an indicator function $ \mathbbm{1}_{C}[c(t)]$ that is one when the contrast $c(t)$ is a small bin around $C$ and zero otherwise.  Then we can generalize Eq~(\ref{STA1}) to give a contrast dependent STA,
\begin{equation}
\label{eq:STAC}
    \STA(\tau; C) = \langle v(t_i - \tau)   \mathbbm{1}_{C}[c(t_i  - \tau)] \rangle .
\end{equation}
One might worry that this definition mixes the response to velocity with the response to time varying contrast, but if we choose the contrast correlation time $\tau_c$ to be longer than the time scales in the STA itself this is not a problem.  We can think instead of $\STA(\tau; C)$ as characterizing a neural response to velocity that is adapted to the local contrast.

As a practical matter we can choose the intervals around $C$ in the indicator function so that we divide the contrast axis into $M$ disjoint bins.  We can choose these to be of fixed size or we can choose them to be equally populated, but since the distribution of $c$ is uniform these are the same up to small sampling errors.  Although we can measure spike arrival times with higher accuracy, our stimulus presentation occurs in steps $\Delta\tau = 2\,{\rm ms}$, and we will measure ${\rm STA}(\tau; C)$ for $|\tau|\leq 200\,{\rm ms}$, corresponding to $L=100$ sampling points.  ${\rm STA}(\tau; C)$ is then an $M\times L$ matrix.  These matrices are the central objects of our analysis.

\subsection{Comparing across distributions}

With the previous section providing a method for estimating how H1 adapts to contrast changes within a contrast distribution, we move to the problem of comparing adaptation across distributions. To begin, consider two slow dynamic contrast variables, $c_1(t) \in \mathcal{C}_1$ and $c_2(t) \in \mathcal{C}_2$, with different instantaneous contrast distributions but the same correlation time $\tau_c$.  Further, let us assume that $\mathcal{C}_1 \subseteq \mathcal{C}_2$. Using the methods described above, we can compute $\STA(\tau, \mathcal{C}_1)$ and $\STA(\tau, \mathcal{C}_2)$ and, if the same number of bins and time-steps are used for each of these estimates, both take the form of an $M \times L$ matrix. Immediately these matrices can be compared qualitatively to visualize  H1's behavior across distributions. Of particular interest is if $\STA(\tau; \mathcal{C}_1 \cap \mathcal{C}_2)$ is identical between the two contrast adapted STAs. If not, this would demonstrate that adaptation to contrast is context dependent and that H1 adapts to contrast at a distributional, rather than instantaneous, level.

\section{Results}

We instantiate these  by looking at the electrical activity of the motion--sensitive neuron H1 in the blue bottle fly {\em Calliphora vicina} (Appendix \ref{app:flies}).  This organism is chosen because it is relatively large, allowing very long stable recordings (Appendix \ref{app:recording}), and because genuinely wild type flies are available in the woods near the laboratory.  The fly visual system is much faster than the human visual system, and we would like to sample the high light intensities that are typical of the mid--afternoon in the natural environment.  To accomplish these goals we use a custom built display system based on an LED array (Appendix \ref{apdx:led_display}).

\subsection{Phenomenology}
\label{sec:phenom}

The central objects in our analysis are the matrices $\STA(\tau; C)$ that describe the sensitivity of the neuron to velocity trajectories when the system is adapted to a particular contrast.  Figure~\ref{fig:stas_clim_60} shows an example of this matrix, plotted as functions of $\tau$ with $C$ as a parameter.  We expect these to be relatively smooth functions, so the high frequency wiggles give a sense for the scale of the errors, and these are consistent with the large but limited number of samples.\footnote{If mean spike rates are $\bar r \sim 20 \, {\rm s}^{-1}$ and we record for $30\,{\rm min}$ then we have $\sim 36,000$ samples with which to compute the spike triggered average.  But these are divided into $M=15$ contrast bins, so we have $N_s \sim 2,400$ samples for each $C$.  The variance of the velocity at a single discrete time point is $\sigma_v^2 = \langle v^2\rangle = \left(37\, {\rm deg/s}\right)^2$, so we expect the estimated $STA$ to fluctuate by $\sigma_v/\sqrt{N_s} \sim 0.75\,{\rm deg/s}$, which is not far off from what we see in Fig~\ref{fig:stas_clim_60}.  The noise is a little larger because not all samples are independent.} The first observation is that the STAs have a width of $\sim 50\,{\rm ms}$ which is an order of magnitude smaller than the correlation time $\tau_c$ for contrast variations, so we have in fact achieved a separation of time scales.

By construction the velocity and contrast are independent of one another, so the different ${\rm STA}(\tau; C)$ represent averages over the same distribution of velocity waveforms.  Dependence of the ${\rm STA}$ on $C$ thus means that the neuron is responding differently to the same inputs, reflecting adaptation to the contrast.  We see that this is an overall change in the amplitude of the ${\rm STA}$, and a change in shape.  Thus contrast drives changes in the sensitivity of the response and the time window over which integration occurs.  Roughly speaking, increasing contrast increases the sensitivity  to velocity and decreases the time over which the neuron integrates.  These changes make sense as adaptations to higher SNR at larger contrast.

\begin{figure}[t]
    \centering
    \includegraphics[width=\linewidth]{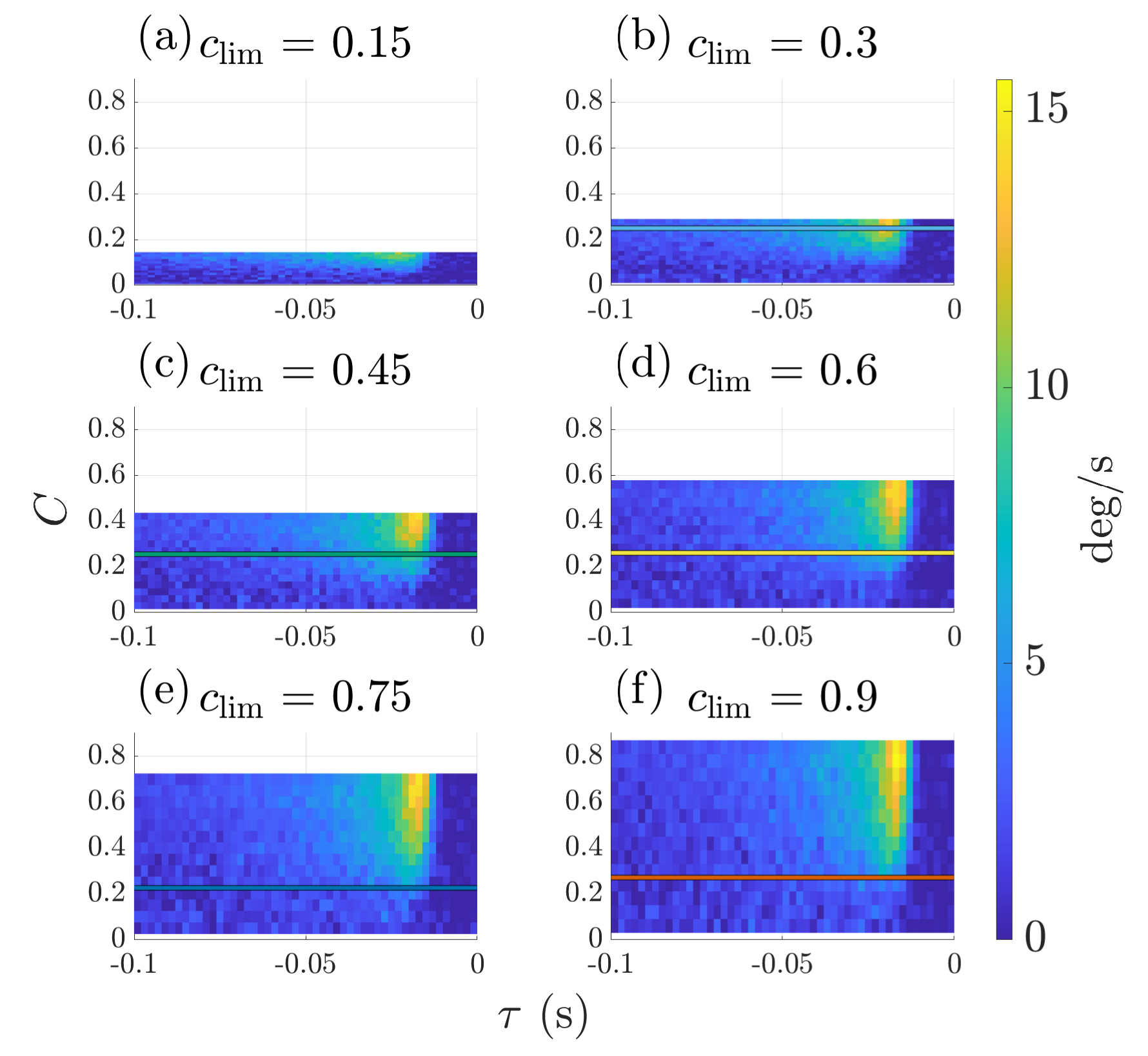}
    \caption{Contrast dependent spike triggered averages, ${\rm STA}(\tau; C)$ from Eq~(\ref{eq:STAC}).  As in Fig.~\ref{fig:stas_clim_60} but for each of the six values of $c_\text{lim}$: (a) $c_\text{lim} = 0.15$, (b) $c_\text{lim} = 0.30$, (c) $c_\text{lim} = 0.45$, (d) $c_\text{lim} = 0.60$, (e) $c_\text{lim} = 0.75$, and (f) $c_\text{lim} = 0.90$. The bold colored lines in (b)-(f) correspond to the contours displayed in Fig.~\ref{fig:stas_at_25percent}.}
    \label{fig:stas_all}
\end{figure}

In our experiments, as in the natural environment, contrast variations are drawn from a probability distribution.  Here this distribution is characterized by its dynamic range $c_{\rm lim}$, and  Fig~\ref{fig:stas_all} shows ${\rm STA}(\tau; C)$ for all values of $\clim$. While line plots were helpful for visualizing STAs within a single distribution, comparing across distributions is more effective when we look at ${\rm STA}(\tau; C)$ more explicitly as matrices.

\begin{figure}[b]
    \centering
    \includegraphics[width=\linewidth]{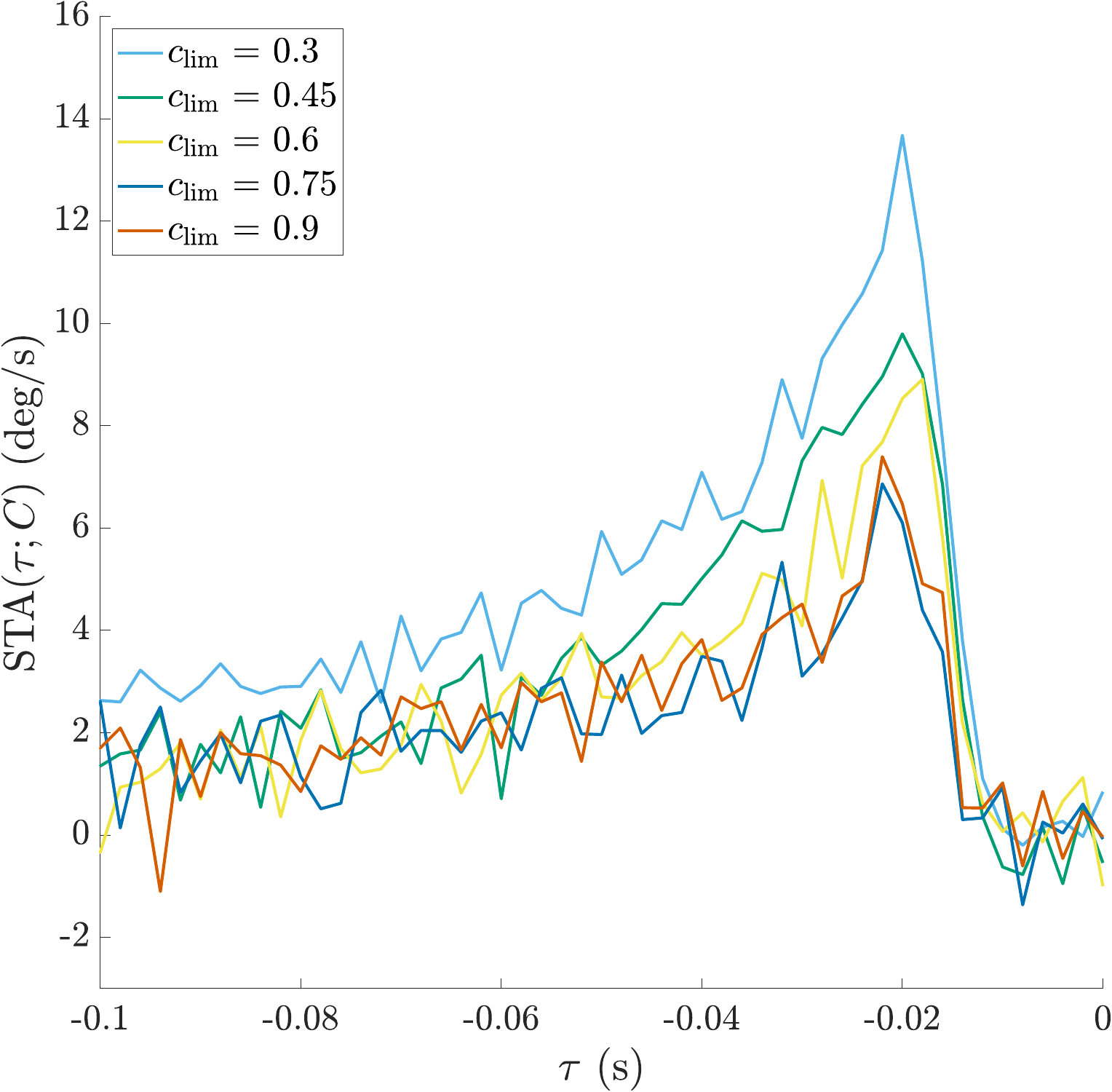}
    \caption{The $\STA(\tau; C = 0.25)$ contour associated with each of the STA surfaces shown in Fig.~\ref{fig:stas_all}. Notice that, despite being the same absolute contrast, the response is different for each $\clim$.}
    \label{fig:stas_at_25percent}
\end{figure}

The STAs are qualitatively similar across distributions, appearing to follow a common   template that is rescaled depending on the contrast distribution. For example, the maximum response magnitude is similar across all distributions and consistently occurs at the highest contrast $C\sim \clim$. This is unexpected---naively, one might have expected STA strength to increase monotonically with contrast, independent of the available dynamic range.  Instead, our result suggests that H1 has a maximum response limit, which it reserves for the highest SNR stimuli. In essence, this shows that contrast adaptation in H1 is context dependent.

Another way to interpret this result is that the response to a given contrast value depends not only on its absolute level but also on the distribution it appears within. This is demonstrated in Fig~\ref{fig:stas_at_25percent}, which shows the ${\rm STA}(\tau; C)$ for $C = 0.25$ across all contrast distributions (excluding $\clim = 0.15$). The responses differ, even though the local contrast and its correlation time are identical in each case.  As the distribution of contrasts becomes broader, the response at a particular contrast decreases, as if adaptation served to normalize the inputs, as happens with the velocity itself \cite{brenner2000adaptive}.

\subsection{A low rank representation}

The smoothness and similarity of the spike triggered averages at different contrasts and in different contexts suggest that these functions occupy only a small part of the high dimensional space of velocity waveforms.  Recalling that each ${\rm STA}(\tau;C)$ can be represented as an $M \times L$ matrix, this is equivalent to saying these matrices do not have full rank.\footnote{For our experiments $M < L$, so that full rank is rank $M$.} This  can be tested for each matrix by computing its singular-value decomposition (SVD) and counting the number of significantly non-zero singular values, which provides our best estimate of the rank $k$.  The right eigenvectors linked to the significant singular values span the dimensions of stimulus space where H1's response adapts, while the left eigenvectors show how the scaling of these basis vectors changes with contrast.  Singular values that are not significant point to dimensions that are dominated by sampling noise, and if we remove these we obtain a smoother rank-$k$ approximation to ${\rm STA}(\tau; C)$.  But this separate analysis of each context misses the possible commonalities across contexts.

\begin{figure}[t]
    \centering
    \includegraphics[width=\linewidth]{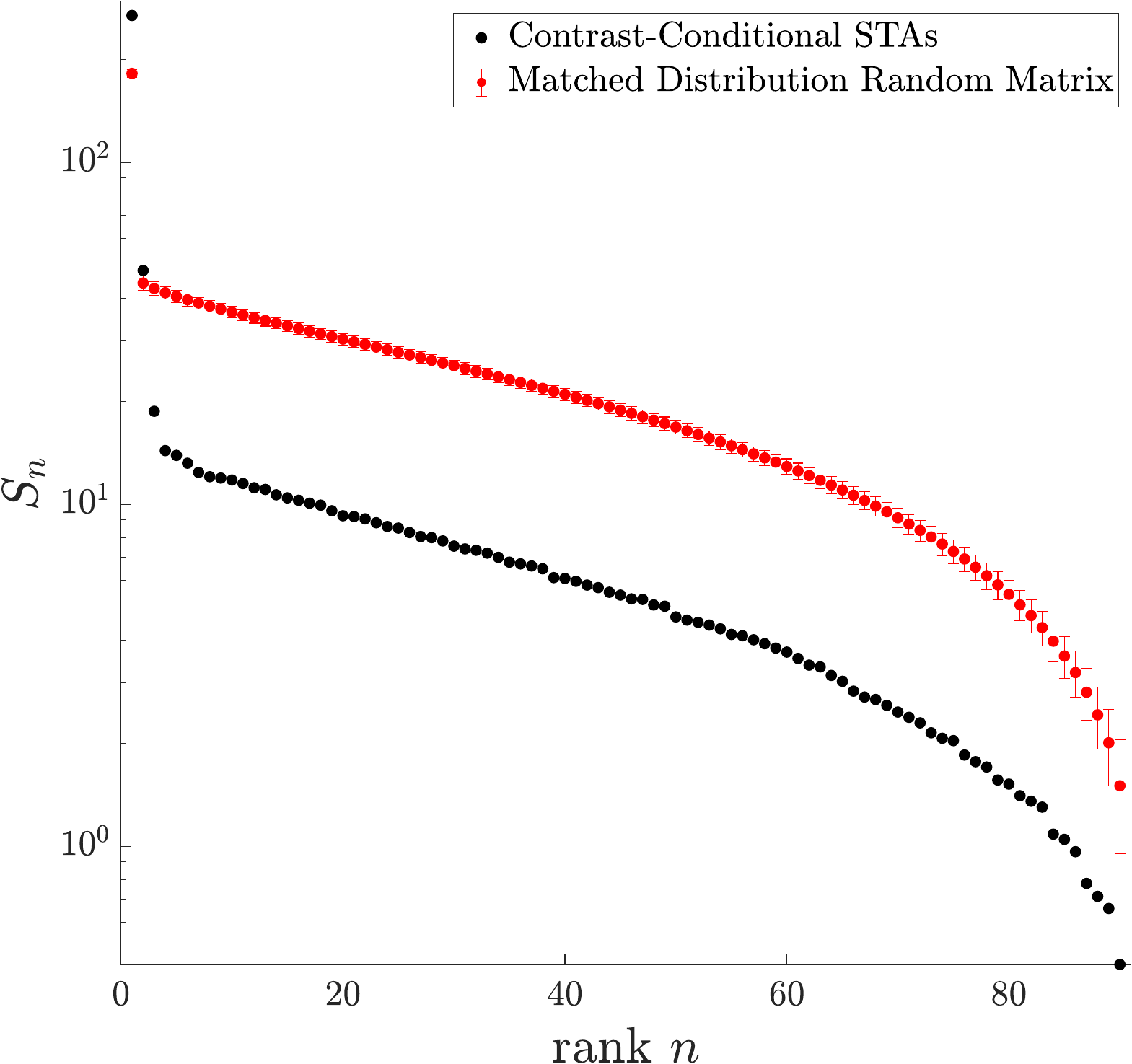}
    \caption{Singular values $S_n$  vs.~rank $n$ for the  spike-triggered averages ${\rm STA}(\tau; C)$ across all six contexts shown  in Fig. \ref{fig:stas_all}a-f. Because we make $M=15$ bins along the $C$ axis and have $L=100$ points along the $\tau$ axis, the full matrix we analyze is $6M\times L = 90\times 100$, so there are at most $90$ independent dimensions.  To test significance we  construct a random $90\times 100$ matrix with elements drawn from the distribution observed in the real data and do the same analysis (red). 
    \label{fig:singular_values}}
\end{figure}

We can try to identify these common structures by forming a new ``stacked" STA, a $6M \times L$ matrix whose rows are indexed by both $C$ and $c_{\rm lim}$. The singular values and right eigenvalues of this matrix maintain their same interpretations while the left eigenvalues are now associated with both contrast and context. Figure \ref{fig:singular_values} shows the singular values that emerge from this analysis of the matrices in  Fig~\ref{fig:stas_all}.   We can test the significance of the singular values by comparing against the behavior of random matrices whose elements are drawn from the distribution of real elements, and we see that just two modes are significant.  This indicates that H1 adapts its sensitivity only  within a two-dimensional subspace of stimulus trajectories.

\begin{figure}[b]
    \centering
    \includegraphics[width=\linewidth]{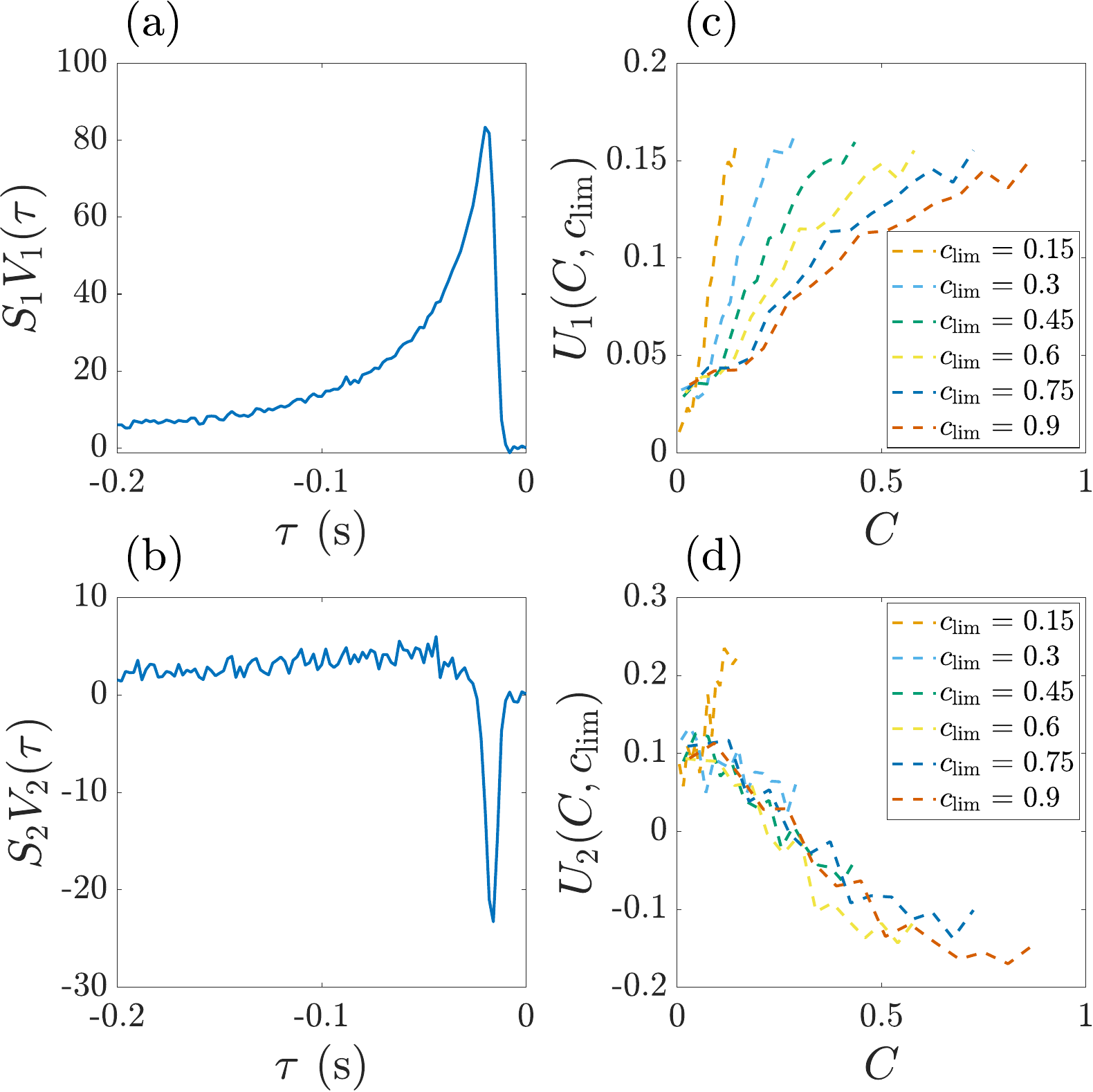}
    \caption{Significant components of the SVD decomposition. (a) The first significant right eigenvector $V_1(\tau)$, shown scaled by its singular value $S_1$.  (b) As with (a), but for $V_2$  (c) The first significant left eigenvector $U_1(C, \clim )$, shown as a function of contrast in different contexts. (d) As with (c), but for $U_2(C; \clim)$. 
    \label{fig:svd_res}}
\end{figure}

The dominance of two dimensions means that we can write, to a good approximation,
\begin{eqnarray}
{\rm STA}(\tau; C, \clim ) &=& U_1(C; \clim) S_1 V_1 (\tau) \nonumber\\
&&\,\,\,\,\,\,\,\,\,\, + U_2(C; \clim) S_2 V_2 (\tau) .
\end{eqnarray}
In this decomposition $V_{1,2}(\tau )$ are normalized basis functions (right eigenvectors of the SVD) in the time domain, $S_{1,2}$ are the singular values that set the overall contribution of these functions to the whole set of STA, and $U_{1,2} (C; \clim )$ provide coordinates (left eigenvectors) in the two dimensional space for each combination of contrast and context.  These functions are shown in Fig.~\ref{fig:svd_res}a--d.

The first velocity eigenvector $V_1(\tau)$, which is associated with the largest singular value, forms the basic profile of the conditional STA while varying contributions from $V_2(\tau)$ serve to shift the peak and adjust its width. Intriguingly, the coordinates or weights $U_{1,2}(C; \clim)$ have very different behaviors as a function of contrast and context.
The weight $U_1(C,\clim)$ has a dependence on contrast that varies strongly with context (Fig.~\ref{fig:svd_res}c), while $U_2(C, \clim)$ depends on contrast but is almost independent of context.

\begin{figure}[t]
    \centering
    \includegraphics[width=\linewidth]{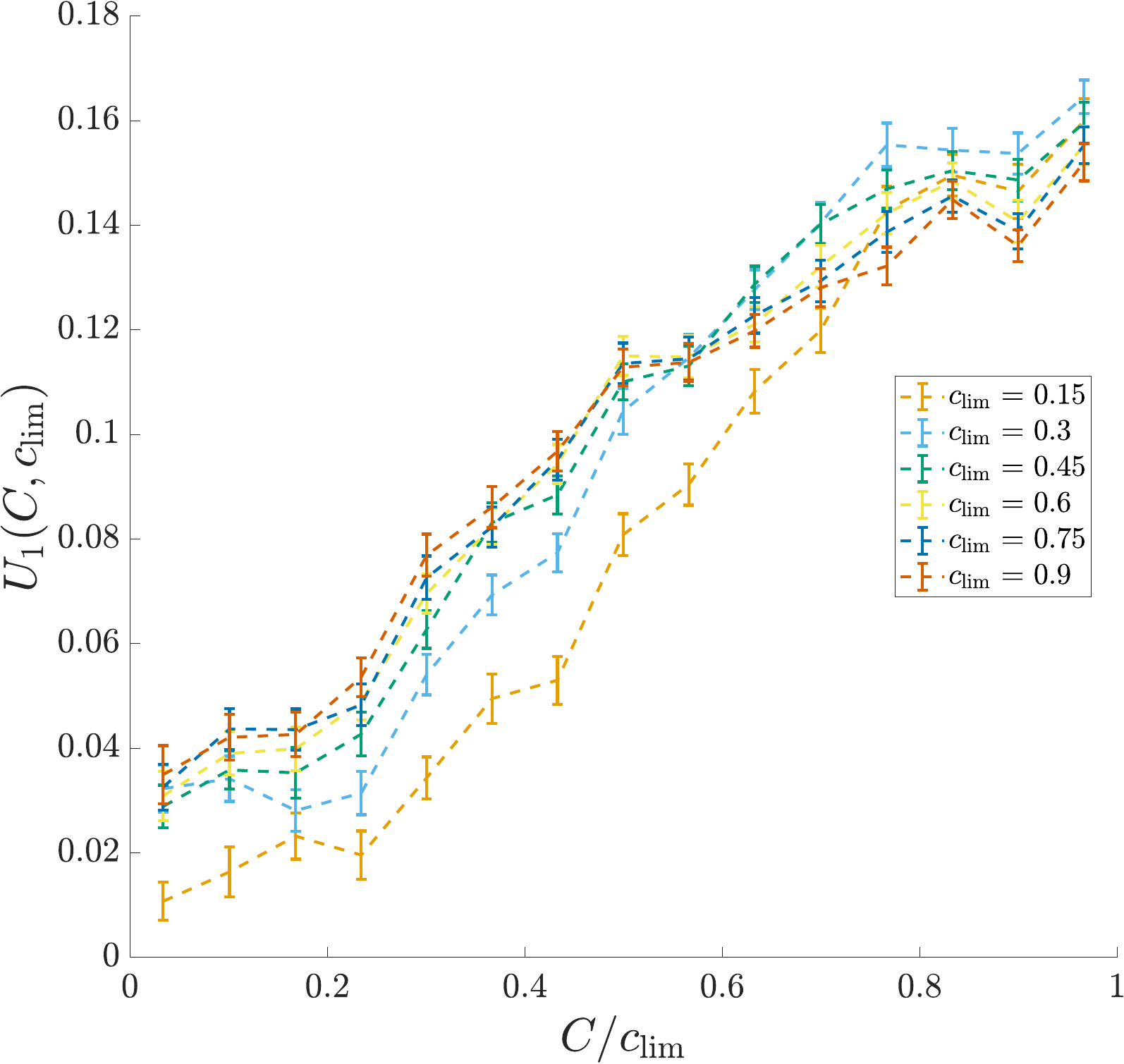}
    \caption{The contribution of the dominant mode to the STA, $U_1(C, \clim )$ from Fig.~\ref{fig:svd_res}c, depends largely on the scaled value $C/\clim$. Deviations are significant only for at the very lowest signal-to-noise ratios, $\clim = 0.15\%$. 
    \label{fig:scaled_eigen}}
\end{figure}

Figure \ref{fig:scaled_eigen} examines the dependence of the $U_1$ on contrast and context more closely. We see that to a very good approximation
\begin{equation}
U_1(C, \clim ) = g(C/\clim),
\end{equation}
so that contrast is normalized to the dynamic range defined by the context.  To be clear, on the $\sim 50\,{\rm ms}$ time scale of the STA,   the movie that the fly actually sees are determined only by $C$, while $\clim$ is something that the system can ``know'' only by accumulating statistics over time scales much longer than $\tau_c = 500\,{\rm ms}$.  We have obtained qualitatively similar results with $\tau_c = 0.25 - 15\,{\rm s}$, indicating that mechanisms of context dependence have access to a wide range of time scales, as seen in the adaptation to the dynamic range of velocity signals \cite{fairhall2001efficiency}.

In summary, the response of H1 to motion depends on both contrast and context.  Surprisingly, this dependence is essentially two dimensional. The more significant dimension varies with contrast scaled to the dynamic range defined by the context, while the second varies with absolute contrast.  Scaling to the dynamic range is the solution to optimal coding problems when the only scale in the problem is provided by the distribution \cite{brenner2000adaptive}.  But this condition is violated if signals are small and noise provides a significant scale, consistent with what we see here at the lowest dynamic range ($\clim = 0.15$).  Such scaling creates an obvious ambiguity, and it is reasonable to think that the second component of the response contributes to this ambiguity being resolvable on longer time scales \cite{fairhall2001efficiency}.

\subsection{Time scales}
\label{sec:timescales}

The diversity of visual systems employing some form of contrast adaptation suggests it serves an important role in optimal information encoding.   In the case of adaptation to the distribution of velocities there is indeed a link between the precise form of the adaptation and the information carried by single spikes or intervals \cite{brenner2000adaptive}; if we look just after a switch in the distribution we can catch the system before it adapts fully and see that information transmission is reduced \cite{fairhall2001efficiency}.  Despite these precedents we were unable to   link the observed contrast and context adaptations to increases in information carried by single spikes.   But because the neuron responds to velocity as seen through the filter provided by the STA, if the width of this filter is too large then successive spikes carry information about overlapping segments of the velocity trajectory.  If the signal-to-noise ratio is low this could help with averaging, but if the SNR is high it would create redundancy.

If the fly's visual system is in a regime where reducing redundancy is a dominant consideration, then an optimal code would insure that successive spikes represent essentially independent pieces of the signal $v(t)$.  This requires that the integration time that we observe through the spike triggered be comparable to or smaller than the typical interval between spikes. To test this hypothesis we need to extract an effective integration time from the STAs at each combination of contrast and context.

As discussed in Appendix \ref{app:timescales}, all the STAs can be fit reasonably well as exponential decays with a delay,
\begin{equation}
{\rm STA}_{\rm fit}(-\tau) = a_1\Theta (\tau - \tau_{\rm delay}) \exp\left[ - (\tau - \tau_{\rm delay})/\tau_{\rm int} \right] + a_2 .
\label{fitform}
\end{equation}
Here the delay is $\tau_{\rm delay}$ and the time constant of the exponential decay $\tau_{\rm int}$ provides an estimate of the integration time; $a_1$ is the overall amplitude of the response and $a_2$ is a very small background that improves the quality of the fit but otherwise has little effect. Details of fitting and error estimates are in Appendix \ref{app:timescales}.

Figure \ref{fig:sta_width_vs_rate} shows the integration time $\tau_{\rm int}$ for all combinations of contrast and context plotted vs.~the corresponding mean spike rate $\bar r$. We see that the data cluster tightly around $  {\bar r} \tau_{\rm int} = 1$, with almost no points significantly above this line.  The system can arrive at the same spike rate with different combinations of $C$ and $\clim$, and the integration time $\tau_{\rm int}\sim 1/\bar r$ in each case. These results suggest that integration times adapt to be as long as possible, reducing noise as much as possible, without introducing redundancy.  Although the dynamic range of mean spike rates is limited, the relation $  {\bar r} \tau_{\rm int} = 1$ is surprisingly precise.

\begin{figure}
    \centering
    \includegraphics[width=\linewidth]{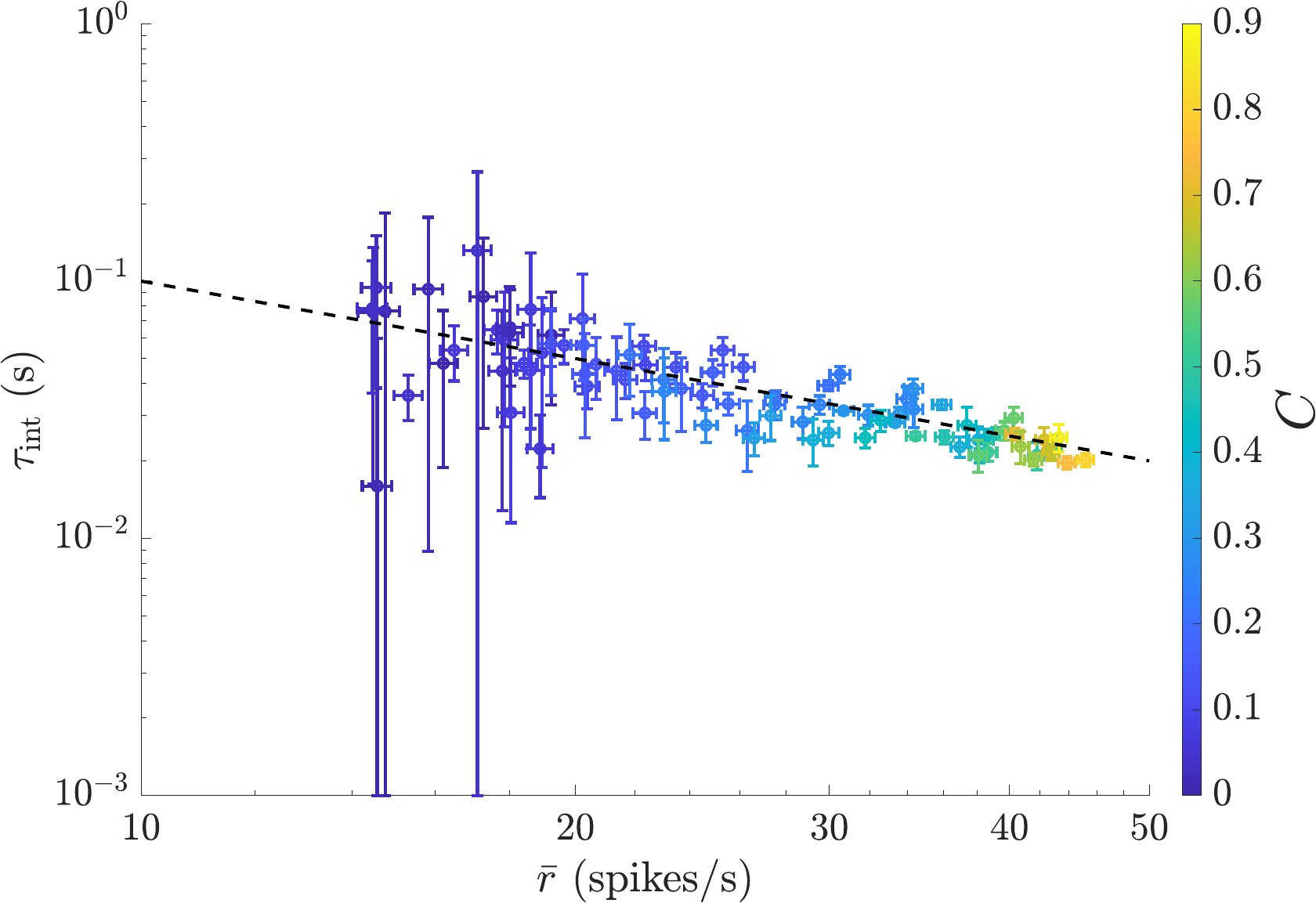}
    \caption{Integration time $\tau_{\rm int}$ of the STA, from Eq.~(\ref{fitform}), at each combination of contrast $C$ (color scale) and context $\clim$, ploted vs.~the corresponding mean spike rate $\bar r$.  Dashed line is $\bar r \tau_{\rm int} = 1$. 
    \label{fig:sta_width_vs_rate}}
\end{figure}

\section{Discussion}

Sensory adaptation sometimes is described colloquially as the brain ignoring uninteresting constant signals.  Barlow's insight was that constant stimuli are not necessarily uninteresting but certainly uninformative, in the technical sense of information theory \cite{barlow_59,barlow1961possible}.  This launched a program of trying to understand adaptation, and neural coding strategies more generally, as solutions to the problem of maximizing information transmission with limited resources. This principle may be more widely applicable to information processing by living systems \cite{bialek2024}.

The motion sensitive neurons of the fly  system, of which H1 is an example, have played an important role in the development of ideas about neural coding and computation.  Although primarily responsive to movement velocity, the contrast dependence of this response is a crucial signature for the algorithm that the brain uses in motion estimation \cite{hassenstein1956systemtheoretische,reichardt1976visual,egelhaaf1989computational,potters1994statistical,van1994statistical,harris2000contrast,sinha2021optimal,borst2023flies}. There is considerable evidence that this contrast dependence is not just (nearly) instantaneous but adaptive \cite{maddess1985adaptation, van1994statistical,de1995reliability, warzecha2000response, harris2000contrast, straw2008contrast}.

In most studies of adaptation the system is exposed for a relatively long time to stimuli with a constant value of the relevant parameter, and then probed with brief transient stimuli.   This experimental design allows the demonstration not only that the ``set point'' of the input/output relation can shift in response to a background but also that the dynamics of responses can change; early experiments on H1 provide an example of this \cite{de1986adaptation}.  An alternative is to look for slow relaxations of the response when the distribution of inputs changes suddenly \cite{smirnakis_adaptation_1997}. But real world signals are not broken into adaptation periods and probes or switches; indeed we know that natural images fluctuate on all length scales, and the local variance or contrast of these images also varies on many scales \cite{ruderman_statistics_1994,ruderman1997origins}. Stimuli with continuously varying local statistics, as constructed here, provided a controlled version of these truly natural inputs \cite{fairhall2001efficiency}.

Our results show that the responses of H1 adapt not just to contrast but to the context provided by the distribution of contrast.  This echoes the adaptation of the same responses to the distribution of velocities \cite{brenner2000adaptive,fairhall2001efficiency}.  In the vertebrate retina we also see adaptation to the dynamic range and spatial scale of contrast variations \cite{smirnakis_adaptation_1997, vinje_natural_2002}, as well as to the distribution of color contrast \cite{vasserman_adaptive_2013}.  After initial observations in the H1 and the salamander retina, adaptation to the variance or dynamic range of inputs was observed in systems including visual \cite{felsen_cortical_2005}, auditory \cite{cooke+al_2018}, and somatosensory \cite{maravall+al_2007} cortices.  

We usually think of light and dark adaptation, for example, as shifting the responses of a neuron to changes in light intensity.  Here we demonstrate that adaptation to contrast and its dynamic range modulate the response of a motion sensitive neuron to its primary stimulus, image velocity.  While such adaptation across dimensions is known, it is not well characterized. Renewed attention to the ``mixed selectivity'' of neurons throughout the brain \cite{tye+al_2024} suggests there may be more opportunities to observe these effects.  

Of all the ways in which the neural response to velocity could vary, we find that adaptation is described very well by variations in a space of just two dimensions. Similar results have been obtained in the salamander retina, albeit with less naturalistic stimuli \cite{liu_spike-triggered_2015}.  We are particularly struck by the fact that the two dimensions separate so cleanly in how they combine contrast and context: one dimension depends almost entirely on contrast in units of its dynamic range, while the other depends on contrast alone.  As noted above, this may allow the system to achieve efficient coding in response to rapid variations while still leaving a smaller signal that resolves ambiguities on longer time scales.

We have emphasized a phenomenological description of adaptation.  In general it can be challenging to provide a clear functionality for the behavior that is observed in such experiments.  We believe, however, that the balancing of spike rates and integration times (Fig.~\ref{fig:sta_width_vs_rate}) provides an important clue.  We note that for many years it was assumed that neurons encode sensory information by generating many action potentials for each characteristic time of the stimulus, so that a local rate of spiking is well defined.  Early measurements on H1 and other systems showed that, instead, neurons often generate of order one spike per characteristic time \cite{spikesbook}.  In this regime one can think of each spike as pointing to some definite event in the sensory world \cite{deruyter+bialek_1988,bialek+zee_1990,bialek+al1991,spikesbook}. This regime is very efficient, in the sense that the information carried by a long sequence of spikes is very nearly the sum of the information carried by each spike, with little redundancy.  The results in Fig.~\ref{fig:sta_width_vs_rate} suggest, strongly, that the system adapts to actively maintain itself in this regime as the context for neural coding changes.

In summary, we have provided evidence that responses of the visual neuron H1 adapt both to local contrast and to the context provided by the dynamic range of contrast variations.  This requires adaptation mechanisms that span a wide range of time scales.  Additionally, we observe that H1 adjusts its contrast-conditional velocity response by modifying within a low-rank subspace, and that this adaptive behavior is dominated by a scaling such that the range of responses is stretched or compressed to place its highest response at the distribution's maximum contrast.  Finally, we saw hints that these adaptations may serve as a method for balancing the trade-off between producing independent spikes and ensuring that each spike carries a significant amount of sensory information. Taken together, our results contribute to the growing body of evidence that neural responses adapt, quantitatively,  to the distribution of sensory inputs, a key feature of efficiently coding systems. 

\begin{acknowledgments}

The authors thank Gary Wood for helping to construct the experimental apparatus, in particular the electrical/light shielding box. We also thank Mike Hosek and Shiva Sinha for their excellent work developing the high-intensity display used for stimulus presentations. This research was supported in part by Lilly Endowment, Inc., through the Indiana University Pervasive Technology Institute, and by Indiana University. Additional support was provided by the National Science Foundation, through the Center for the Physics of Biological Function (PHY--1734030), and by fellowships from the Simons Foundation and the John Simon Guggenheim Memorial Foundation (WB). 
\end{acknowledgments}

\appendix

\section{Fly Husbandry}
\label{app:flies}

All recordings were conducted using either wild-caught or lab-bred wild-type blue bottle flies (\textit{Calliphora vicina}). The wild-caught flies were collected from Dunn Woods at Indiana University Bloomington and housed in a clear enclosure inside an opaque cabinet. They were provided with water, protein powder, and sugar cubes. LED lights on a timer maintained a 12 hour day/night cycle and a passive humidifier maintained relative humidity levels above $40 \% $ year round. For breeding, fly eggs were collected by placing chicken liver in the fly enclosure for 24 hours. The eggs were then raised until pupation in a specialized ``baby hotel"---a second cabinet with no light source and increased ventilation---before the pupated flies were transferred to a new enclosure within the fly housing cabinet. Only F0, F1, and F2 generation flies were used in the experiments.

\section{Neural recording}
\label{app:recording}

To prepare a fly for recording, an active, uninjured fly was first selected and removed from the enclosure by hand. The fly's wings were restrained with a small drop of melted dental wax, and the fly was immobilized in a small plastic cylinder so that its head and shoulders protruded above the edge of the retaining tube. Special care was taken to insure that the spiracles were not obstructed and that the fly could freely move its proboscis. The head was gently tilted forward by hand, and wax bridges spanning from the jowls to the propleuron were used to immobilize the head in a forward-tilted position. A small semicircular incision was made in the back right side of the fly's head with a razor blade, and the integument was removed. Excess fat and membranes were cleared to expose the right lobula plate. A dorsiventral-oriented muscle running along the proximal side of the lobula plate was cut to avoid electrical interference during recording. Finally, a small feeding platform was constructed on the top edge of the immobilization tube, just below the fly's proboscis.

The prepared fly was then placed on a tri-axial stage 32 cm in front of the high-intensity LED display, so that the LEDs formed a square raster with angular nearest neighbor separations of 0.91 degs. For comparison, blue bottle fly ommatidia have angular separations of 1.57 deg \cite{beersma1977retinal}. The stage was housed within an aluminum shielding box to block both electrical signals and stray light, preventing interference with the experiments. An FHC dissection scope was used to view the back of the fly's head and two tungsten microelectrodes (5 $\mu$m tip diameter; 1 MΩ resistance) -- a reference and a signal electrode -- were carefully positioned by hand in the fly’s lobula plate to record from the contralateral H1 neuron. The electrode signals were band-pass filtered and differentially amplified using a Princeton Applied Research PAR 113 Pre-Amp.

H1's receptive field was aligned with the display by rotating the tri-axial stage until the response was maximized with respect to a horizontally oscillating checkerboard stimulus and minimized with respect to a vertically oscillating checkerboard stimulus. The positions of the electrodes and the fly’s orientation were adjusted until distinct spikes (signal-to-noise ratio $\geq$ 3) were clearly observed using the horizontal test pattern; see, for example, Fig 1.2 of Ref.~\cite{spikesbook}. After the fly was properly aligned, it was given a sugar water treat before leaving it undisturbed for 5 minutes. This acclimated the fly to the experimental conditions and helped to stabilize recordings. Following this, stimulus presentations and recordings started, capturing H1's spike times along with the frame sync and frame fault signals from the high-intensity display. 

H1 spikes were identified using a World Precision Instruments 121 Window Discriminator and were time-stamped  and recorded at $10~\mu{\rm s}$ sampling resolution using a Cambridge Electronic Design CED Power1401 and SPIKE2 software. Display signals were also recorded at the same time resolution using the same CED Power1401 and SPIKE2 software, ensuring all signal times were recorded in reference to the same clock. Throughout the recording we monitored the raw signals and adjusted the spike threshold to account for any drift in  response amplitude. After the first stimulus ended the recording was saved to disk and the next stimulus was presented. Stimuli were presented from lowest to highest contrast, beginning with $\clim = 0.15$ and ending with $\clim = 0.90$. 

Finally, a single red LED was mounted above the fly and used to illuminate it during the experiment. This LED served two purposes: First, although flies are not spectrally sensitive to red light, they use it in the early visual system to photoregenerate rhodopsin from metarhodopsin \cite{schwemer1983pathways, schwemer1984renewal}. Given the high intensity of the LED display, the red light helped avoid rhodopsin depletion in the fly's photoreceptors \cite{zhou2014dynamics}. Second, it allowed us to observe the experiment without introducing a light source visible to the fly.  All results shown in the main body of this text are from a single large male blue bottle fly (\textit{Calliphora vicina}), however these result were replicated across a total of eight (8) different flies of mixed size and sex.

\section{The High-Intensity LED Display}
\label{apdx:led_display}

Providing the fly visual system with naturalistic stimuli requires a very high frame rate, high absolute intensities, and a wide dynamic range.  To address these issues we developed a custom 48 x 48 high-intensity LED display capable of producing daylight-level light intensities across a 12-bit dynamic range. The display uses 3.0 mm x 2.0 mm  green surface-mount Kingbright AA3021WG1S LEDs, which have a maximum viewing angle of 125 degs and a peak emission of 500 nm at 25 $^\circ$C. The LEDs were arranged into 6 x 4 blocks with 5 mm LED lattice spacings and each block was controlled by an individual Texas Instruments TLC595, which sets LED intensities using 12-bit pulse-width modulation (PWM). The ninety-six (96) blocks were arranged into 12 columns and 8 rows, for an overall display size of 24.2 mm x 24.2 mm, and the TLC595 data lines for blocks in the same column were connected in series, resulting in a total of 12 data channels for all 2304 LEDs. 

Each recording was 30 minutes in length (900051 frames) and all stimuli used the same binary synthetic natural scene and white-noise velocity trace but with variations in $\clim$, as described in Sec.~\ref{sec:control_contrast}. The mean light intensity was fixed at 7 \Wmsr (screen intensity 256),  matching the mid-afternoon light intensities in the local Indiana deciduous hardwood forests where the flies were collected. 

Image data are sent to the display along the 12 parallel data lines using a National Instruments PCIe-6536B digital I/O card and the NI-DAQmx software from a dedicated display computer. Display frame rate and play timing is controlled by a Xilinx Artix-7 FPGA on a Digilent Cmod A7 development board. The display has a maximum display rate of 500 Hz and play timing can be set using one of two methods: an internal 500 Hz clock provided by the Digilent Cmod A7 or an external clock signal. Additionally, the FPGA generates three digital timing signals: a frame signal, a sync signal, and a fault signal. The frame signal goes high 100 $\mu$s before a frame is displayed. The sync signal goes high for the first two display frames, and then every 100 display frames thereafter. The fault signal goes high only if a frame fails to load before the next play signal is received. Together, these signals provide high-quality timing information about when a frame was, or was not, displayed. All experiments used a 500 Hz play rate with play signals generated by the CED Power1401's clock. The frame, sync, and fault signals all were timestamped using the same clock as the H1 recording channel, as noted above.

To obtain a quantitative measurement of the display's intensities, LED brightness was recorded using the FlEye camera \cite{edelson+al_2024} during a linear ramp of PWM intensities. These measurements were then used to generate a linear response curve relating PWM values to radiance. The best-fit linear coefficient from the ordinary least squares (OLS) analysis was $0.028\,{\rm Wm}^{-2}{\rm sr}^{-1}$ per PWM unit. This means that a PWM intensity of 256 corresponds to a radiance of 7$\,{\rm Wm}^{-2}{\rm sr}^{-1}$. This equates to an effective photon rate $\sim 10^6\,{\rm s}^{-1}$ per  photoreceptor in the fly retina, which is similar to the photon rates a fly experiences in bright daylight conditions.

\section{Rates and integration times}
\label{app:timescales}

As explained in Section \ref{sec:timescales}, we would like to compare the spike rates of H1 with the integration times for velocity signals. To extract the rate as a function of contrast we follow the same strategy as with the $\STA$, using an indicator function to select those moments in time the contrast is within small bin surrounding the value $C$.  Then we can estimate the probability of a spike given the contrast
\begin{equation}
    \hat{p}(C) = \frac{\sum_{n,i} \delta(t_n - t_i) \mathbbm{1}_C[c(t_n)]}{\sum_n \mathbbm{1}_C [c(t_n)]},
\end{equation}
where we emphasize that the times $t_n = n \cdot \Delta \tau$ in our experiment are clocked in discrete steps of duration  $\Delta \tau = 2$ ms. The estimated contrast-conditional average spike rate is then 
\begin{equation}
    \hat{\bar{r}}(C) = \frac{\hat{p}(C)}{\Delta \tau}.
\end{equation}
We can associate with each such measurement a standard error of the mean (SEM) that comes just from the binary counting of spikes, 
\begin{equation}
    \hat{{\rm SEM}} = \sqrt{\frac{\hat{p}(C)(1 - \hat{p}(C))}{T \cdot \Delta \tau}},
\end{equation}
with $T$ the total duration of the experiment,
and these are the error bars on $\bar r$ in Fig.~\ref{fig:sta_width_vs_rate}. Very similar error estimates were also produced from a bootstrap procedure where we look at random fractions of the data.  This agreement presumably reflects the fact that spikes are separated by intervals comparable to the integration time (Fig.~\ref{fig:sta_width_vs_rate}), so that variability in different time bins is almost independent.

\begin{figure}[t]
    \includegraphics[width=\linewidth]{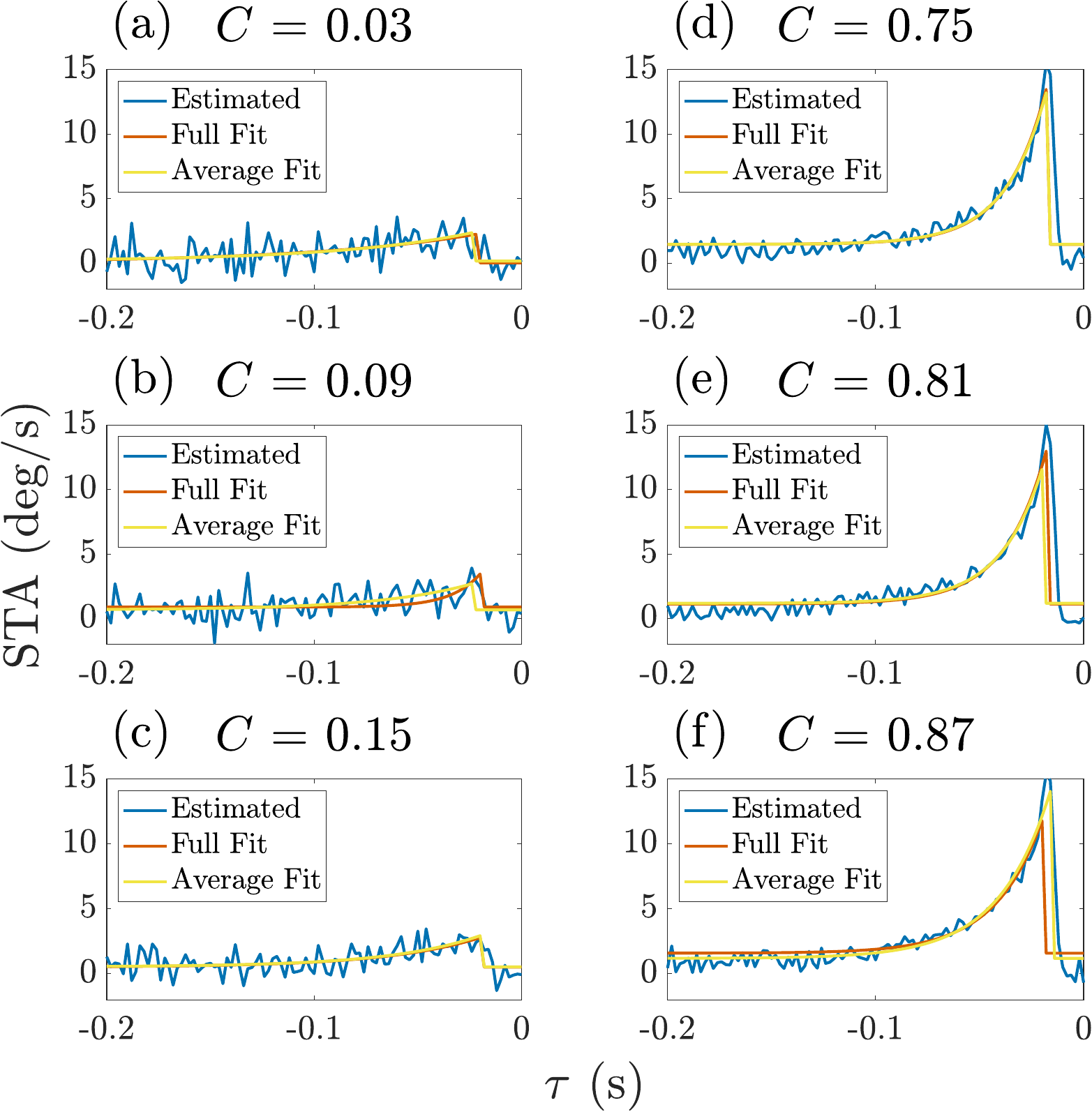}
    \caption{Example of contrast-conditional STAs and their exponential fit for $\clim = 0.9$ using Eq.~\ref{fitform} and Eq.~\ref{eq:loss} for contrast of (a) 0.03, (b) 0.09, (c) 0.15, (d) 0.75, (e) 0.81, and (f) 0.87. Estimated STAs are shown in blue, while the fit produced using all available data is shown in red. Moreover, the fit produced by the mean fit parameters is shown in yellow. Not that while both fits are generally similar, there are a few exceptions. For example, the average fit for (b) appears to better capture the wider shape of this low contrast STA.
    \label{fig:fit_example}}
\end{figure}

To estimate the STA integration time $\tau_{\rm int}$, we first computed the contrast-conditional STAs for each recording as described in Sec.~\ref{sec:in_out_and_corr}. At high contrast, these STAs resemble a time-shifted exponential decay (Fig.~\ref{fig:stas_clim_60}). We leveraged this property by fitting each contrast-conditional STA to Eq.~(\ref{fitform}), minimizing the mean square error
\begin{equation}
\label{eq:loss}
    \mathcal{L} = \int dt\, {\bigg |} \STA(t; C) -  \STA_{\rm fit}(t;a_1, a_2,\tau_{\rm delay}, \tau_{\rm int}) {\bigg |}^2 .
\end{equation}
Fits were constructed using MATLAB's \texttt{fminsearch} function, a numerical non-linear minimum finder that uses the simplex search method   \cite{lagarias1998convergence}; examples  for $\clim = 0.9$ are shown in Fig.~\ref{fig:fit_example} in red. It is apparent that the fit curve nicely follows the functional form of the contrast-conditional STA even at low contrast levels.

Given the non-linear fitting procedure used above, these is no closed form representation of the errors in these fits, as we have with a simpler statistic such as $\bar{r}$. To remedy this, a data-dropping and refitting method was used to estimate both the average fit parameter and fit parameter variances for the $\STA$ at each contrast and context.  This was done by randomly masking 10\% of the 2 ms stimulus time intervals in the recording before refitting the contrast-conditional STA with the 90\% data-set. Next, a new set of coefficients were fit to this ``resampled" contrast-conditional STA. This process was repeated 64 times, creating a resampled empirical distribution of parameter values, from which we estimated the mean and standard deviation. Examples of fits produced using the average fit parameter are shown in Fig.~\ref{fig:fit_example} in yellow. 

\bibliography{dissertation_bibliography}

\end{document}